\documentclass[twocolumn,showpacs,preprintnumbers,amsmath,amssymb,prb]{revtex4}

\usepackage{graphicx,graphics}
\usepackage{bm}

\begin{document}
\title{Thermal rectification in
carbon nanotube intramolecular junctions: Molecular dynamics
calculations }

\author{Gang Wu}
\email{wugaxp@gmail.com} \affiliation{Department of Physics and
Centre for Computational Science and Engineering, National
University of Singapore, Singapore 117542-76, Republic of Singapore
}

\author {Baowen Li}
\email[Corresponding author. Electronic address:
]{phylibw@nus.edu.sg} \affiliation{Department of Physics and Centre
for Computational Science and Engineering, National University of
Singapore, Singapore 117542-76, Republic of Singapore }
\affiliation{ NUS Graduate School for Integrative Sciences and
Engineering, Singapore 117597, Republic of Singapore}

\begin{abstract}
We study heat conduction in ($n$, 0)/(2$n$, 0) intramolecular
junctions by using molecular dynamics method.  It is found that the
heat conduction is asymmetric, namely, heat transports preferably in
one direction. This phenomenon is also called thermal rectification.
The rectification is weakly dependent on the detailed structure of
connection part, but is strongly dependent on the temperature
gradient. We also study the effect of the tube radius and
intramolecular junction length on the rectification. Our study shows
that the tensile stress can increase rectification. The physical
mechanism of the rectification is explained.

\end{abstract}


\date{\today}
\maketitle

\section{Introduction}

In past two decades, the study of heat conduction in low dimensional
model systems has enriched our understanding on heat conduction from
microscopic point of view\cite{r2}. In turn, the study has also lead
to some interesting inventions for heat control and management
devices. For example, the heat conduction in nonlinear lattice
models \cite{r3, r4,JHLan06} demonstrates rectification phenomenon,
namely, heat flux can flow preferably in one direction. Furthermore,
the \textit{negative differential thermal resistance} is also found
and based on which, a thermal transistor model has been
constructed.\cite{r5} Most recently, the two segment model of
thermal rectifier proposed in Ref \onlinecite{r3, r4} has been
experimentally realized by using gradual mass-loaded carbon and
boron nitride nanotubes.\cite{r6} These works have opened a new era
for heat management and heat control in microscopic level.

On the other hand, there have been increasing studies on heat
conduction in real nano scale systems\cite{r1}. For example, the
thermal conduction of single walled carbon nanotubes (SWCNTs) has
attracted both theoretical \cite{r12, r13, r14, r15, r16, r17, r18,
r19, r20, r21} and experimental \cite{r22, r23, r24, r25, r26, r27,
r28} attentions. Almost all experiments and numerical simulations
have payed their attention to the extremely high thermal
conductivity of SWCNTs. The dependence of the thermal conductivity
on the length or chirality of SWCNTs is also extensively studied
theoretically. Therefore, one may asks, how can we make heat
controlling devices, i.e., the thermal diode or the thermal
transistor, from SWCNTs and their derivatives?

In the practical applications, SWCNTs may have many kinds of
impurities such as element impurities, isotopic
impurities\cite{ZLi05}, topological impurities, etc. Among these
numerous derivatives of SWCNTs, the SWCNT intramolecular junctions
(IMJs), which are formed by introducing the pentagon-heptagon rings
in SWCNTs, have been expected to be an important structure in the
applications. It is well-known that the electronic properties of the
SWCNT IMJs have a close relationship with their geometrical or
topological characteristics.\cite{r29, r30, r31, r32, r33, r34, r35,
r36} It is reasonable that the thermal transport behavior of SWCNT
IMJs can also be altered by their geometrical characteristics.
Compared with the electronic properties, the thermal transport
behavior may have not so close relationship with the detailed local
geometrical arrangements because it is mainly influenced by the long
wavelength phonons. But the non-equilibrium thermal transport
behavior of IMJs can be affected by the high-frequency optical
phonon modes,\cite{r37} which can reflect the information of the
local defects. So it is interesting to investigate the
non-equilibrium thermal transport of different SWCNT IMJs.

In this work, the thermal rectification in ($n$, 0)/(2$n$, 0) IMJs
and its dependencies on tube radius, IMJ length, and external
stress, e.g., tensile and torsional stress are studied. A complex
structure of `peapod' structure and IMJ is also studied to
investigate the effect of periodical potential on the thermal
rectification.

The paper is organized as the follows. In Sec. II we introduce the
basic structure of (n, 0)/(2n, 0) IMJ and our numerical method. In
Sec. III, the main numerical results are discussed. In Sec IV, we
show that the external stress can improve the rectification. In the
same section we also show the results of (n, 0)/C$_{60}$@(2n, 0)
structures, which can be regarded as a combination of peapod
structures and (n, 0)/(2n, 0) IMJs. In Sec. V, we give some
concluding remarks.

\section{Model and Methodology}

A typical ($n$, 0)/(2$n$, 0) IMJ structure is depicted in Fig. 1, in
which the index $n$ equals to 8. The structure contains two parts,
namely, a segment of ($n$, 0) SWCNT and a segment of (2$n$, 0)
SWCNT. For simplicity, the lengths of the two segments are almost
equal in our calculations. The two segments are connected by $m$
pairs of pentagon-heptagon defects. Because ($n$, 0) and (2$n$, 0)
tubes have the common rotational symmetry $C_{2n}$, thus when $n$
can be divided exactly by $m$, we can adjust the defects to make the
two segments connect straightly. Especially, if $m = n$, the
pentagon-heptagon defects can be arranged in the connection part so
that the rotational symmetry of the IMJ is $C_{n}$. On the other
hand, it is necessary to define a basic system length because the
thermal conductivity depends on the system length.\cite{ZLi05} After
setting the lengths of ($n$, 0) and (2$n$, 0) tubes to be the same,
we can define a basic total length $L_0 $ when both of the two
segments contain 24 periods. All the structures are fully optimized
before further molecular dynamic (MD) calculations, and then we can
obtain $L_0 \approx 20$ nm.

\begin{figure}[htbp]
\includegraphics[height=\columnwidth,angle=-90]{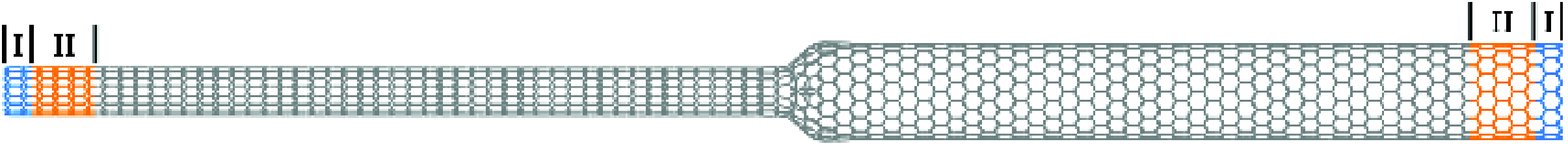}
\label{fig1} \caption{(Color online). A typical ($n$, 0)/(2$n$, 0)
structure. Here, $n = 8$, and the number of pentagon-heptagon
defects is $m = 4$. The regions marked as `I' are fixed in MD
process. The regions marked as `II' are put in the heat baths.}
\end{figure}

In this kind of structure, the outmost one period of each heads
(colored by blue and marked as region `I' in Fig. 1) are fixed in MD
process. Then two periods of each ends (illustrated by orange color
and marked as region `II' in Fig. 1) are put in the heat baths,
which are realized by the Nos\'e-Hoover thermostat.\cite{NoseHoover}
The temperatures of the thermostats at left and right heads are $T_L
$ and $T_R $, respectively. For convenient, here we introduce two
quantities: $\Delta T = \frac{T_L - T_R }{2}$, and $\left\langle T
\right\rangle = \frac{T_L + T_R }{2}$. In this work, $\left\langle T
\right\rangle $ is always kept at 290 K.

The C-C bonding interactions are described by the
second-generation reactive empirical bond order (REBO)
potential,\cite{r38} which is the most recent version of the
Tersoff-Brenner type potential, combining advantages of the two
sets of parameters in the earlier version.\cite{r39}

The velocity Verlet method is employed to integrate the equations
of motion with the time step of 0.51 fs. The typical total MD
process is 5$\times $10$^{6}$ steps, which is about 2.55 ns, and
the statistic averages of interesting quantities start from half
of the MD process, i.e., 2.5$\times $10$^{6}$ steps are used to
relax the system to a stationary state.

The instant temperature of atoms $i$ is defined as $T_i \left( t
\right) = \frac{m_i }{3k_b }\left( {v_x \left( t \right)^2 + v_y
\left( t \right)^2 + v_z \left( t \right)^2} \right)$, where
$v(t)$ is the time-dependent velocity, $m_i $ the mass and $k_b $
the Boltzmann constant. This is a result of energy equipartition
theorem.

The thermal flux is obtained from the thermostats, i.e., the total
work from the thermostats can be regarded as the heat flux runs from
thermostats to the system. If the work is positive, then the heat
flux is also positive. In the scheme of the Nos\'e-Hoover
thermostat,\cite{NoseHoover} the equation of motion for the particle
$i$ in heat bath is:

\begin{equation}
\label{eq1} {\dot{\vec p}_i}=- \xi {\vec p_i}+{\vec f_i}, \quad
\dot{\xi}=\frac{1}{Q}\left( \frac{{\vec p_i}\cdot {\vec p_i}}{m_i} -
g k_b T \right),
\end{equation}

\noindent where $\vec p_i$ is the momentum and $\vec f_i$ is the
force applied on the atom. $g$ is the number of degrees of freedom
of the atoms in the thermostat. $Q=g k_b T {\tau }^2$, where $\tau$
is the relaxation time. In our simulation, $\tau $ is kept as 4 ps.
The heat bath acts on the particle with a force $ - \xi {\vec p_i}$,
thus the power of heat bath is $ - \xi \frac{{\vec p_i}\cdot {\vec
p_i}}{m_i}$, which can also be regarded as the heat flux from heat
bath, i.e.,

\begin{equation}
\label{eq2} J_i = - \xi \frac{{\vec p_i}\cdot {\vec p_i}}{m_i}.
\end{equation}

The total heat flux from the heat bath to the system can be
obtained by $J=\sum _i J_i$, where the subscript $i$ runs over all
the particles in the thermostat. The final thermal conduction is
$\kappa \cdot s = \frac{j \cdot s}{2\Delta T / L} =
\frac{JL}{2\Delta T }$, where $j$ is the heat flux density, $s$
the area of cross-section, and $L$ the system length. In fact this
definition is equivalent to the usual definition: $J =
\frac{d}{dt}\sum\limits_i {{\vec r} _i \left( t \right)\varepsilon
_i \left( t \right)} $, $\varepsilon _i \left( t \right)$ is the
instant total energy of particle $i$, but Eq. (\ref{eq2}) is much
simpler for computational simulation.

\section{Numerical results and discussions}

First of all, we investigate the effect of defects in connection
region. The defects in connection region determine the property of
the interface between two tubes, and as a result, they will affect
the heat transport property of the system. However, the question
whether they can affect the rectification is still open.

We study the (8, 0)/(16, 0) structures as examples.

Generally speaking, they are many methods to connect an IMJ, thus
we only consider two conditions here. The first one is the coaxal
straight IMJs. The possible number $m$ of pentagon-heptagon defect
pairs for a coaxal IMJ is 8, 4 or 2, but $m = 2$ makes the
cross-section of the tube part highly deformed from circle, so we
only consider $m = 4$ and 8. The second condition is the simplest
IMJ, i.e., two tubes are connected with each other by using only
one pair of heptagon-pentagon defects ($m = 1$).\cite{SaitoBook}
The top views and side view of the three structures ($m = 8$, 4
and 1) are plotted in Fig. 2. It can be found even in the simplest
zigzag/zigzag IMJ ($m=1$), there is still a small angle between
the two segments.

\begin{figure}[htbp]
\includegraphics[height=\columnwidth, angle=-90]{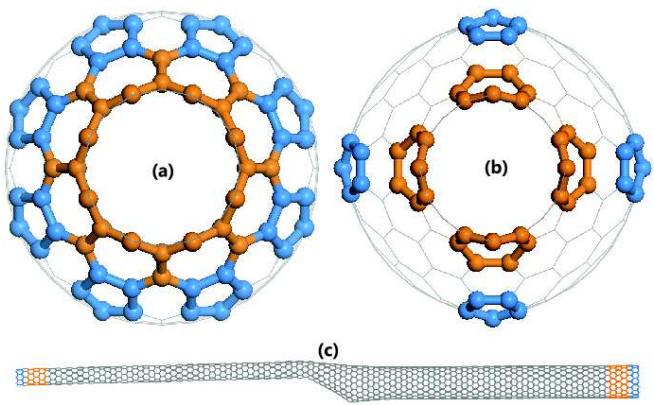}
\label{fig2}
\caption{(Color online). The top views and side view of two (8,
0)/(16, 0) IMJs with different connection methods. (a) With 8 pairs
of pentagon-heptagon defects. (b) With 4 pairs of pentagon-heptagon
defects. (c) With only one pair of pentagon-heptagon defects.}
\end{figure}

The total heat fluxes $J$ versus the normalized temperature
difference ${\Delta T} / {\left\langle T \right\rangle }$ is plotted
in Fig. 3a. Positive $J$ means the heat flux flows from ($n$, 0)
tube to (2$n$, 0) tube. We do not show the heat flux density here,
because it is difficult to define the area of the cross-section of
the whole IMJ.

\begin{figure}[htbp]
\includegraphics[width=0.8\columnwidth]{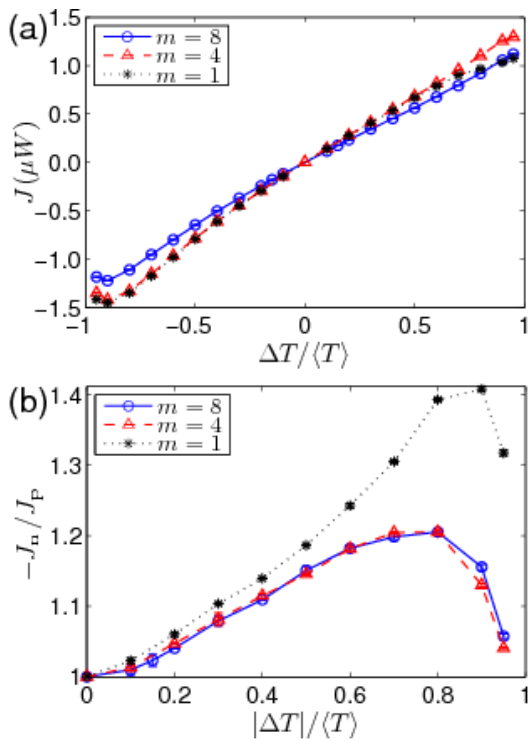}
\label{fig3} \caption{(Color online). The thermal transports of (8,
0)/(16, 0) SWCNT IMJs with different numbers of pentagon-heptagon
defects in the connection region. (a) The heat flux versus
temperature difference. (b) The thermal rectification versus
temperature difference. $m$ is the number of pentagon-heptagon
defects. Error bars are also plotted.}
\end{figure}

It can be found from Fig. 3a that the total heat flux for $m=4$ is
larger than that for $m=8$. It is straightforward because the
pentagon-heptagon defects are topological defects, they scatter
the phonons and cause larger interfacial thermal resistance. As a
result, more topological defects will reduce the thermal
conductance and the heat flux. But the total heat flux for $m=1$
is close to that for $m=4$. This is due to the big difference of
the connecting region of these IMJs. The length of the connecting
region in IMJ with $m = 1 $ is much longer than those in the IMJs
with $m = 4$ or 8. The long connecting region can cause big
thermal resistance, as a result, the thermal resistance in $m=1$
IMJ is not much smaller than that in $m=4$ IMJ.

When the temperature difference is very small, the heat flux only
shows weak asymmetry. The asymmetry becomes larger when
temperature difference is larger. This characteristic is seen
clearer in Fig. 3b, in which the thermal rectification
$|J_n/J_p|=-J_n/J_p$ is drawn.

In Fig. 3b, $J_ n $ means that the heat flux is of negative sign,
\textit{vice versa}, $J_ p$ means that the heat flux is of
positive sign. It can be seen that, an increase in temperature
gradient $|\Delta T|$ leads to an increase of thermal
rectification. This is because the rectification is a result of
non-equilibrium transport, which is mainly determined by optical
phonons. Only when the temperature difference is large enough, the
optical phonons can be excited and contribute to the heat
conduction. More interestingly, when the temperature difference is
large enough (for $|\Delta T| \approx 0.8{\langle T \rangle }$),
the rectification begins to decrease. But, the temperature
difference is so large that this result may not be hold in more
restrict calculations due to lose of local thermal equilibrium and
the quantum effect in the lower temperature head. Another
meaningful feature is that, although the absolute values of the
heat fluxes are completely different, rectifications are almost
the same in IMJs with $m=4$ and 8, whose connecting region is
short. Namely, the rectification is weakly dependent on the
detailed structure of the interface when the connecting region is
short enough. And the rectification in $m=1$ IMJ is largest among
the three kinds IMJs.

In order to understand above rectification phenomenon,  we
consider the power spectra of the atoms around the connecting
part. The atoms to be studied are illustrated in the Fig. 4a. They
are selected to be next to the defects along the axes. The
corresponding phonon spectra are presented in the Fig. 4b. And the
overlapping of the phonon spectra of the two atoms around the
connecting parts have been emphasized by shadows.

\begin{figure}[htbp]
\includegraphics[width=0.65\columnwidth]{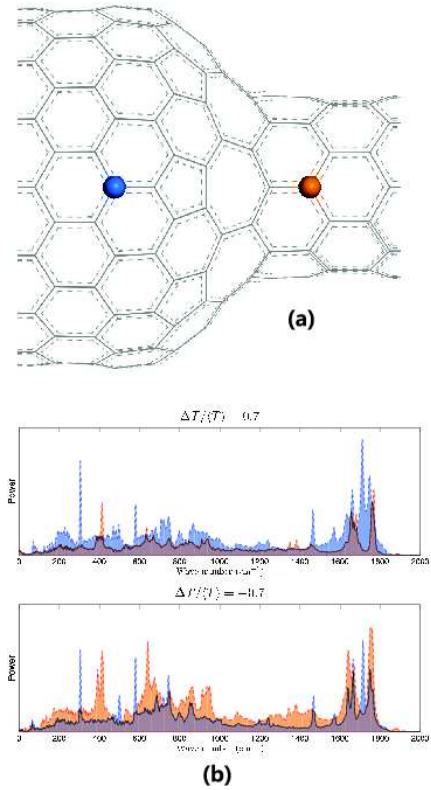}
\label{fig4} \caption{(Color online). (a) The side view of the (8,
0)/(16, 0) IMJ ($m=8$), where the atoms whose power spectra are
recorded are emphasized by balls. (b) The corresponding phonon
power spectra. The overlapping of the phonon spectra of the two
particles around the connecting parts have been emphasized by
shadows and bordered by dark lines.}
\end{figure}

 It is found that when the direction of the
temperature gradient is exchanged, the area of the overlap region
is also changed. More importantly, the overlap area is larger when
${\Delta T} /{\langle T \rangle }$ is negative. This result
corresponds with the trend of the heat flux. So we can state that
in this real system, the relationship between the overlap area and
the absolute value of the heat flux is same as that in the one
dimensional nonlinear lattice systems\cite{r3, r4, JHLan06}, in
which it is found that matching/mismatching of the energy spectra
around the interface is the underlying mechanism of the
rectification.

On the other hand, the most obvious change appears at 400$\sim
$1000 cm$^{ - 1}$ and 1600$\sim $1800 cm$^{ - 1}$. The $m=4$ IMJ
is also studied, and similar conclusion can be obtained. So the
optical phonon modes are important for the rectification. In
addition, in Ref. \onlinecite{r37}, the authors show that in
carbon nanotubes with finite length where the long-wavelength
acoustic phonons behave ballistically, even optical phonons can
play a major role in the non-Fourier heat conduction. The
dispersion relations of the SWNT show that, in the intermediate
range of the normalized wave vector $0.1 < k^ * < 0.9$, some of
the phonon branches, especially the ones with relatively low
frequency, have group velocity comparable to the acoustic
branches.

Furthermore, the thermal conductivity can be expressed as $\kappa
= \sum\limits_q {l _{\lambda q} C_q \left( {\omega _\lambda }
\right)v_{\lambda q} } $, where $\lambda $ is a set of quantum
numbers specifying a phonon state, $l $ is the phonon mean free
path, $v _{\lambda q}$ is the magnitude of the phonon group
velocity along the direction of the heat flow, and $C_q \left(
\omega \right) = k_B \left( {\frac{\hbar \omega }{k_B T}}
\right)^2\frac{\exp \left( {{\hbar \omega } /{k_B T}}
\right)}{\left[ {\exp \left( {{\hbar \omega }/{k_B T}} \right) -
1} \right]^2}$ is the thermal capacity of lattice wave with wave
vector $q$ and angular frequency $\omega $. Thus, the contribution
from the high frequency modes is weak because their group
velocities are much smaller than those of low frequency modes. So
the modes at 400$\sim $1000 cm$^{ - 1}$ are more important than
the modes at 1600$\sim $1800 cm$^{ - 1}$.

The frequencies of the low frequency modes have been found to be
inversely proportional to the tube radius.\cite{Saito1998} Thus,
the low frequency modes, e.g., the radial breathing mode, have
different exciting temperatures in (16, 0) and (8, 0) tubes. On
the other words, their participation in transport process occurs
at different temperatures. This fact further explains why these
low frequency optical phonon modes are important for the
rectification.

Based on above discussions, now we can understand the following
phenomena. Firstly, the rectification almost does not change in
the $m=4$ and 8 IMJs. Secondly, the rectification in $m=1$ IMJ is
largest in the three IMJs.

As has been mentioned, the modes at 400$\sim $1000 cm$^{- 1}$ are
most important for the thermal rectification. These modes have
larger wave lengths than the high frequency modes. The connecting
parts of the two IMJs with $m = 4$ and 8 are short, as a result,
the defects have weak effect on the effective optical modes. In
Ref. \onlinecite{Wu2006}, it is found that the high frequency
vibrational modes caused by topological defects are localized
states. And the higher frequency localized modes show smaller
spatial dimension, which means they can hardly entangle with the
transport phonon modes. Thus in the short connecting IMJs ($m = 4$
and 8), the arrangement of the detailed defects has weak influence
on the rectification. In contrast, the connecting region is much
longer in the $m = 1$ IMJ, and the distance between the pentagon
and heptagon rings is relatively long, which means the connecting
region can affect those middle frequency phonon modes, which have
longer wave lengths. Therefore, the rectification is largest in
the case of $m = 1$ IMJ.

However, the rectification in current structures is still small,
which might limit the application of the IMJ as a thermal
rectifier. So we will try to find the factors which can help
improve the rectification.

Above all, we investigate the thermal conductance
of the IMJs with the different index $n$ numerically.

\begin{figure}[htbp]
\includegraphics[width=0.8\columnwidth]{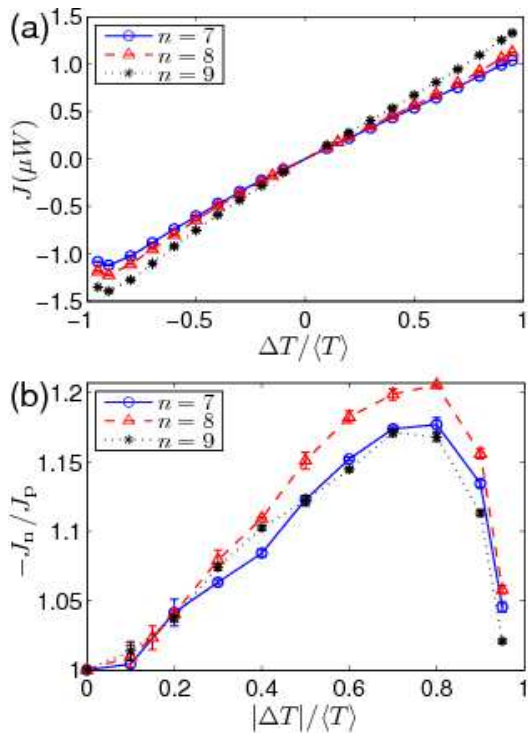}
\label{fig5} \caption{(Color online). The thermal transports of
($n$, 0)/(2$n$, 0) SWCNT IMJs, $n$ = 7, 8 and 9. (a) The heat flux
versus temperature difference. (b) The thermal rectification versus
temperature difference. In this figure, $m = n$. Error bars are also
plotted.}
\end{figure}

In order to make the results comparable with each other, the number
$m$ of pentagon-heptagon defect pairs are set to be $n$. Considering
the computational consumption, only $n = 7$, 8 and 9 are considered
in Fig. 5.

According to Fig. 5a, the total heat flux increases when $n$
increases, but this does not mean that the thermal conductivity
for big $n$ is also large because it should be scaled by the area
of cross-section. In fact, it has been shown that the thinner
SWCNTs have higher thermal conductivity in literatures. \cite{r18,
r21, r26}

Another interesting feature is that, when index $n$ increases, the
asymmetric heat flux ratio does not always increase. In Fig. 5b,
the largest rectification appears when $n = 8$. So this reminds us
that the heat flux rectification is not only induced by the radius
difference of two segments. When tube radius increases to a
sufficient large value, i.e., $n$ is large, the vibrational
density of states (VDOS) of carbon nanotube approaches to that of
the graphite. But when $n$ is small, the VDOS of the SWNT deviates
from that of the graphite due to the periodical boundary along the
circumferential direction and the curvature effect, and the
amplitude of the deviation is almost inversely proportion to $n$.
As a result, when ($n$, 0) and (2$n$, 0) tubes connect with each
other, the difference between the VDOSs of the two tubes is
obvious if $n$ is small, and \emph{vice versa}, the difference is
not so obvious if $n$ is large. Recalling the fact that the
rectification can be related with the mismatching between the
VDOSs around the connecting region, we can conclude that the
rectification of the IMJ is not obvious when $n$ is large.

\begin{figure}[htbp]
\includegraphics[width=0.8\columnwidth]{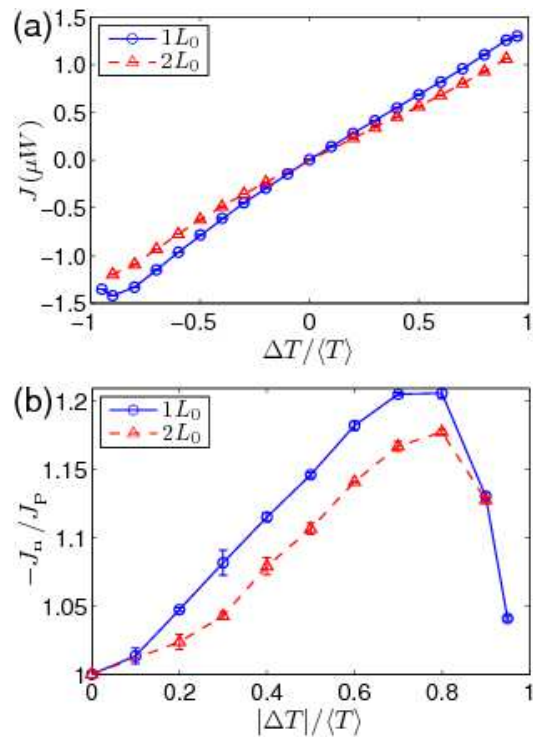}
\label{fig6}\caption{(Color online). The thermal transports of (8,
0)/(16, 0) SWCNT IMJs with different length. (a) The heat flux
versus temperature difference. (b) The thermal rectification versus
temperature difference. $m = 4$. $L_0 $ is defined in text. Error
bars are also plotted.}
\end{figure}

Next, we investigate the length dependence of heat conduction. From
Fig. 6a, it can be seen that the total heat flux decreases as the
system length increases from $L_0 $ to $2L_0 $. However, the thermal
conductivity, $\kappa = \frac{JL}{2s\Delta T}$, increases with the
increasing of the system length. This result agrees with the
previous studies. Moreover, there still exists asymmetry in the heat
flux, which is clearly illustrated in Fig. 6b. Where the
rectification at different $\left| {\Delta T} \right|$ is plotted
for different system lengths. Obviously, the increase of system
length weakens rectification. It can be understood as follows. As
the system length increases, long-wave phonons modes contribute more
to the heat conduction. This kind of heat conduction is symmetric
because long-wave phonon modes are hardly scattered by the junction
or any local defects. Moreover, the asymmetric thermal conductance
is mainly controlled by the asymmetric interface, i.e., the
connection part, which does not change when system length increases.
As a result, although the absolute value does not change, the
rectification appears to be smaller.

Based on above study, we can conclude that changes in radial and
axial direction have different influence on rectification. The
increase of structural asymmetry around the interface can increase
the rectification. In the following section we will discuss other
alternatives to improve the rectification.

\section{Improvement}

First, it is possible to increase the difference in force constants
around the interface by applying external stress. Because the
Young's modulus has been proved to be almost the same in different
tubes using the Tersoff-Brenner potential,\cite{r40} it can be
expected that the ($n$, 0) tube and (2$n$, 0) tube will give
different response to the same external stress. Thus it is possible
that one can use this property to fabricate devices. Here we
consider two kinds of stresses, tensile and torsional stresses. It
has been proven in the framework of tight-binding \cite{r41} that
these two stresses can change the electronic structure of SWCNTs
greatly. Here the (8, 0)/(16, 0) SWCNT IMJ ($m = 4)$ is taken as an
example. The original length is $L_0 $.

The stresses considered in this work are all small enough to ensure
the system does not undergo any structural transformations \cite{r42}
or defects.\cite{r43} For tensile stress, we elongate the IMJ along
the axial direction by 3{\%}. For torsional stress, one head of IMJ
is rotated relative to the other by 30\r{ } around the axis. The
results are shown in Fig. 7.

\begin{figure}[htbp]
\includegraphics[width=0.8\columnwidth]{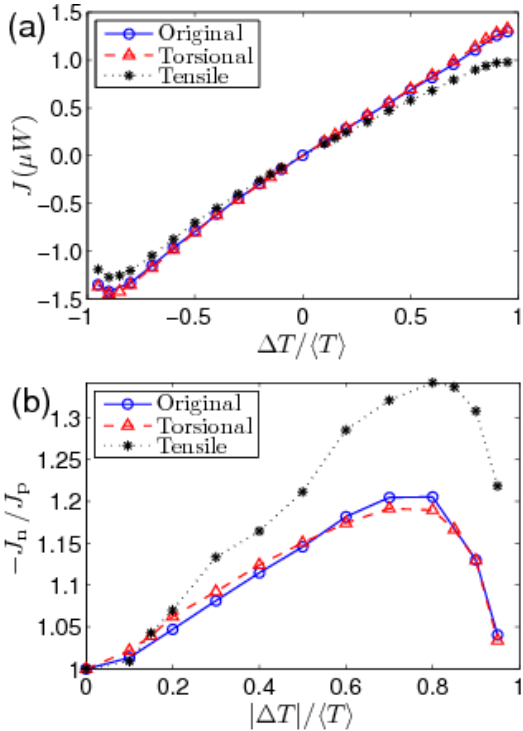}
\label{fig7} \caption{(Color online). The thermal transports of (8,
0)/(16, 0) SWCNT IMJs under different stresses. (a) The heat flux
for different stresses. (b) The thermal rectification for different
stresses. $m = 4$. Error bars are also plotted.}
\end{figure}

It can be seen that the torsional deformation can change neither the
absolute value of heat flux nor the asymmetric behavior largely.
However, the tensile stress can greatly change the heat transport.
In the elongated structure, the heat flux becomes smaller than that
of undeformed structure, but the rectification becomes larger.

Above results are reasonable. First of all, the deformation of the
thicker tube is smaller than that of thinner ones under same external
stress. This means that when thick and thin tubes are connected with
each other to form IMJ, they will appear different axial deformation
under the same stress, which will further increase the structural difference
between the two tubes. Secondly, the heat transport in carbon nanotubes are
mainly contributed by the four acoustic phonon modes. These phonon
modes have weak dependence on the local structure, but they can be
affected by the shape of the cylinder surface. When tensile stress
is applied to the SWCNTs, the SWCNTs will shrink in the radial
direction and elongate in the axial direction. This means their
cylinder surface will change, so that the acoustic phonon modes will
also be changed. But when SWCNTs are twisted, the cylinder surface
almost does not change, so that the thermal conductance almost does
not change.

Comparing Fig. 7a and 3a, it can be found that the heat flux decreases in elongated structure. The reason is straightforward. As mentioned above, the thermal conductivity can be expressed as $\kappa = \sum\limits_q {l _{\lambda q} C_q \left( {\omega _\lambda } \right)v_{\lambda q} } $, where $C_q \left( \omega \right) = k_B \left( {\frac{\hbar \omega }{k_B T}} \right)^2\frac{\exp \left( {{\hbar \omega } /{k_B T}} \right)}{\left[ {\exp \left( {{\hbar \omega }/{k_B T}} \right) - 1} \right]^2}$. For the low frequency ($\hbar \omega \ll k_B T)$ modes, which are most important for the thermal transport, $C_q \left( \omega \right) \approx k_B $. Furthermore, $l_{\lambda q} $ and $v_{\lambda q} $ are the increasing functions of the force constants $k$, thus the thermal conductivity $\kappa $ is also an increasing function of the force constants $k$. In fact, some numerical simulations show that $\kappa \propto k^2$ in the weak coupled lattice models.\cite{r3, WeakLink} When the system is elongated, the bonds along the axis are also elongated. Then according to the Abell-Tersoff formalism of the Tersoff-Brenner potential, the corresponding force constants will decrease. Finally, the thermal conductivity decreases in the elongated structure.

To elongate an IMJ is not very difficult for the state-of-the-art experiments, so we believe that tensile stress can be a possible method to improve the rectification.

Next, we will try to modify the structure to change rectification.

It is suggested \cite{r3,r4} that a suitable on-site potential can
induce large rectification in two-segment model. So it is quite
interesting that what will happen if some periodical potential is
introduced into the SWCNT IMJs. On the other hand, a nanotube with
sufficiently large diameter can be filled with spherical C$_{60}$
fullerene molecules to build up a new hybrid structure referred to
as a `peapod',\cite{r45, r46} with spherical fullerenes representing
peas and the carbon nanotube representing a pod. The thermal
conductivity of the peapod structure has been investigated by using
MD recently.\cite{r47} So it is possible to combine these two
structures, i.e., IMJs and peapod, into a new structure.

\begin{figure}[htbp]
\includegraphics[height=\columnwidth,angle=-90]{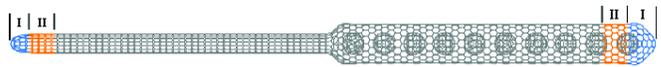}
\label{fig8} \caption{(Color online). The (8, 0)/C$_{60}$@(16, 0)
structure. The number of pentagon-heptagon defects is $m = 4$. The
regions marked as `I' are fixed in MD process. The regions marked as
`II' are put in the heat baths.}
\end{figure}

The structure (see Fig. 8) contains two parts, i.e., a segment of
($n$, 0) SWCNT and a segment of (2$n$, 0) SWCNT with some C$_{60}$
balls inserted into its center, which is typically called as SWCNT
peapod structure. The lengths of these two segments are almost the
same. They are connected by $m$ pairs of pentagon-heptagon defects.
Because the closest distances between the adjacent C$_{60}$ balls
are almost same,  10.05 {\AA}, we can consider that they exert an
external periodic potential on the outside SWCNT. Here, the
interaction between two atoms on C$_{60}$ balls and outside SWCNT is
modeled by Lennard-Jones potential:

\begin{equation}
\label{eq3} V\left( r \right) = 4\varepsilon \left[ { - \left(
{\frac{\sigma }{r}} \right)^6 + \left( {\frac{\sigma }{r}}
\right)^{12}} \right],
\end{equation}
with parameters, $\varepsilon = 2.964$ meV and $\sigma = 3.407$
{\AA}. By default, the total length of the structure is $L_0 $,
which permits 10 C$_{60}$ balls in the (2$n$, 0) segment.

In this complex structure, the two heads are closed to ensure the
C$_{60}$ balls not running out in the MD calculations. And several
periods of two heads (presented by blue color and marked as region
`I' in Fig. 8) are fixed in the MD simulation. Then two periods of
each ends (emphasized by orange color and marked as region `II' in
Fig. 8) are put in the Nos\'e-Hoover thermostat.\cite{NoseHoover}

In Fig. 9, the heat transports of (8, 0)/C$_{60}$@(16, 0) structure
with different lengths are shown.

\begin{figure}[htbp]
\includegraphics[width=0.8\columnwidth]{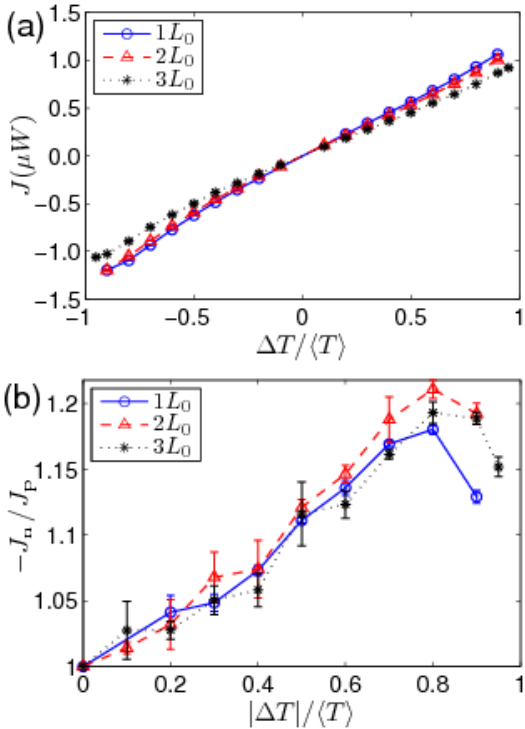}
\label{fig9} \caption{(Color online). The thermal transports of (8,
0)/C$_{60}$@(16, 0) structures with different length. (a) The heat
flux versus temperature difference. (b) The thermal rectification
versus temperature difference. $m = 4$. $L_0 $ is defined in text.
Error bars are also plotted.}
\end{figure}

Comparing Fig. 9a with Fig. 6a, one can find that the heat flux of
peapod structure is smaller than that of the empty IMJ, which seems
to be in contradict with the result in Ref. \onlinecite{r47}. But it
should be pointed out that in our structure the peapod region is
closed at the ends. In fact, this prevents the possible mass
transport (via fullerenes) which in turn reduces the assistance
transport effect of C$_{60}$ balls. Furthermore, because of the
coupling between the phonon modes (mainly the radial breathing
modes) of carbon nanotube (CNT) and C$_{60}$ balls, the conduction
phonon will be scattered and the thermal conductivity of outside CNT
will be reduced.

On the other hand, we can notice that the length of error bars in
Fig. 9b is much larger than that in Fig. 6b. This means that the
existence of C$_{60}$ balls increases the fluctuations in the heat
flux. In Ref. \onlinecite{r47}, it is found that the temperature
fluctuations of peapod structure are larger than that of empty
nanotube. It is explained by the weak van der Waals forces between
fullerenes and the outside CNT.

Another result is that the existence of C$_{60}$ balls can hardly
change the asymmetric behavior of heat flux. This is somehow
disappointing for us, because it is natural to expect that the
periodical potential introduced by the C$_{60}$ balls can enhance
the thermal rectification according to the results based on lattice
models. However, our numerical results are also reasonable because
of following facts. The periodical potential introduced by the
C$_{60}$ balls is intrinsically van der Waals potential, so its
strength is much weaker than the binding potential among the carbon
atoms in SWCNT. The period of the periodical potential introduced by
C$_{60}$ balls is about 10 {\AA}, which is too large compared with
the C-C bond length in SWCNT, thus only low frequency vibrational
phonons are affected. As has been mentioned, the asymmetric heat
flux is controlled by the optical phonon modes, so the change of low
frequency vibrational modes can hardly change the asymmetric
behavior.

According to our result, if one wants to build a thermal rectifier by applying periodical external potential, van der Waals force is not a good choice.

\section{Conclusions}

In this work, heat rectification in ($n$, 0)/(2$n$, 0) IMJs has been studied. It is found that the rectification depends weakly on the detailed structure of connection part, but depends strongly on the temperature gradient. The rectification increases as the temperature gradient increases.

We have also studied the dependencies of the rectification on tube radius and IMJ length. We found that the maximum rectification appears in short (8, 0)/(16, 0) IMJ.

Moreover, we have found that the tensile stress can increase the rectification, while torsional stress can not. It is due to the different effects of the stresses on IMJ structure. It can be believed that the tensile stress will play an important role in building of devices with highly asymmetric transport behavior.

The `peapod' structure is combined with IMJ structure to investigate the effect of periodical potential on the rectification. Our numerical result shows that the new structure can hardly change the rectification.

The study may shed lights in further experimental investigation in this field.

\begin{acknowledgments}
This work is supported in part by an academic research fund of
MOE, Singapore, and the DSTA under Project Agreement No.
POD0410553.
\end{acknowledgments}


\begin{thebibliography}{99}

\bibitem {r2} F. Bonetto, J. L. Lebowitz, and L. Rey-Bellet, In \textit{Mathematical Physics 2000}, edited by A. Fokas, A. Grigoryan, T. Kibble, and B. Zegarlinsky (Imperial College Press, London, 2000) pp. 128-150; S. Lepri, R. Livi, and A. Politi, Phys. Rep. \textbf{377}, 1 (2003); B. Li, J. Wang, L. Wang, and G. Zhang, Chaos \textbf{15}, 015121 (2005).

\bibitem {r3} B. Li, L. Wang, and G. Casati, Phys. Rev. Lett. \textbf{93}, 184301 (2004).

\bibitem {r4} B. Li, J.H. Lan, and L. Wang, Phys. Rev. Lett. \textbf{95}, 104302 (2005).

\bibitem{JHLan06} J. H. Lan and B. Li, Phys. Rev. B \textbf{74}, 214305
(2006); J. H. Lan and B. Li, Phys. Rev. B \textbf{75}, 214302
(2007).

\bibitem{NYang07} N. Yang, N.- B Li, L Wang, and B Li, Phys. Rev. B \textbf{76}, 020301(R) (2007).

\bibitem {r5} B. Li, L. Wang, and G. Casati, Appl. Phys. Lett. \textbf{88}, 143501 (2006).

\bibitem {r6} C. W. Chang, D. Okawa, A. Majumdar, and A. Zettl, Science \textbf{314}, 1121 (2006).

\bibitem {r1} D. G. Cahill, W. K. Ford, K. E. Goodson, G. D. Mahan, A. Majumdar, H. J. Maris, R. Merlin, and S. R. Phillpot, J. Appl. Phys. \textbf{93}, 793(2003).

\bibitem {r12} J.W. Che, T. Cagin, W.A. Goddard, Nanotechnology \textbf{11}, 65 (2000).

\bibitem {r13} S. Maruyama, Physica B \textbf{323}, 193 (2002); S. Maruyama, Microscale Thermophysical Engineering \textbf{7}, 41 (2003).

\bibitem {r14} W. Zhang, Z.Y. Zhu, F. Wang, T.T. Wang, L.T. Sun, and Z.X. Wang, Nanotechnology \textbf{15}, 936 (2004).

\bibitem {r15} N.G. Mensah, G. Nkrumah, S.Y. Mensah, and F.K.A. Allotey, Phys. Lett. A \textbf{329}, 369 (2004).

\bibitem {r16} Z. Yao, J.S. Wang, B. Li, and G.R. Liu, Phys. Rev. B \textbf{71}, 085417 (2005).

\bibitem {r17} N. Mingo and D.A. Broido, Nano Lett. \textbf{5}, 1221 (2005); N. Mingo and D.A. Broido, Phys. Rev. Lett. \textbf{95}, 096105 (2005).

\bibitem {r18} G. Zhang and B. Li, J. Phys. Chem. B \textbf{109}, 23823 (2005).

\bibitem {r19} M. Grujicic, G. Cao, and W.N. Roy, J. Materials Science \textbf{40}, 1943 (2005).

\bibitem {r20} J. S. Wang, J. Wang and N. Zeng, Phys. Rev. B \textbf{74}, 033408 (2006).

\bibitem {r21} J. Wang and J.S. Wang, Appl. Phys. Lett. \textbf{88}, 111909 (2006).

\bibitem {r22} J. Hone, M. Whitney, C. Piskoti, and A. Zettl, Phys. Rev. B \textbf{59}, R2514 (1999).

\bibitem {r23} S. Berber, Y.K. Kwon, and D. Tomanek, Phys. Rev. Lett. \textbf{84}, 4613 (2000).

\bibitem {r24} P. Kim, L. Shi, A. Majumdar, and P.L. McEuen, Phys. Rev. Lett. \textbf{87}, 215502 (2001).

\bibitem {r25} C.H. Yu, L. Shi, Z. Yao, D.Y. Li, and A. Majumdar, Nano Lett. \textbf{5}, 1842 (2005).

\bibitem {r26} M. Fujii, X. Zhang, H.Q. Xie, H. Ago, K. Takahashi, T. Ikuta, H. Abe, and T. Shimizu, Phys. Rev. Lett. \textbf{95}, 065502 (2005).

\bibitem {r27} E. Pop, D. Mann, Q. Wang, K. Goodson, and H.J. Dai, Nano Lett. \textbf{6}, 96 (2006).

\bibitem {r28} T. Y. Choi, D. Poulikakos, J. Tharian, and U. Sennhauser, Nano Lett. \textbf{6}, 1589 (2006).

\bibitem{ZLi05} G. Zhang and B. Li, J. Chem. Phys,\textbf{123}, 014705(2005); C. W. Chang, A. M. Fennimore, A. Afanasiev, D. Okawa, T. Ikuno, H. Garcia, Deyu Li, A. Majumdar, and A. Zettl, Phys. Rev. Lett. \textbf{97}, 085901 (2006).

\bibitem {r29} B. I. Dunlap, Phys. Rev. B \textbf{49}, 5643 (1994).

\bibitem {r30} J.-C. Charlier, T.W. Ebbesen, and Ph. Lambin, Phys. Rev. B \textbf{53}, 11 108 (1996).

\bibitem {r31} L. Chico, V. H. Crespi, L. X. Benedict, S. G. Louie, and M. L. Cohen, Phys. Rev. Lett. \textbf{76}, 971 (1996).

\bibitem {r32} R. Saito, G. Dresselhaus, and M. S. Dresselhaus, Phys. Rev. B \textbf{53}, 2044 (1996).

\bibitem {r33} V. Meunier, P. Senet, and Ph. Lambin, Phys. Rev. B \textbf{60}, 7792 (1999).

\bibitem {r34} M. S. Ferreira, T. G. Dargam, R. B. Muniz, and A. Latg\'{e}, Phys. Rev. B \textbf{62}, 16040 (2000).

\bibitem {r35} L.F. Yang, J.W. Chen, H.T. Yang, and J.M. Dong, Euro. Phys. J. B \textbf{33 (2)}, 215 (2003).

\bibitem {r36} J. Han, M. P. Anantram, R. L. Jaffe, J. Kong and H. Dai, Phys. Rev. B \textbf{57}, 14983 (1998).

\bibitem {r37} J. Shiomi and S. Maruyama, Phys. Rev. B \textbf{73}, 205420 (2006).

\bibitem {NoseHoover} S. Nose, J. Chem. Phys. \textbf{81}, 511 (1984); W. G. Hoover, Phys. Rev. A \textbf{31}, 1695 (1985).

\bibitem {r38} D.W. Brenner, O.A. Shenderova, J. A. Harrison, S. J. Stuart, B. Ni, and S. B. Sinnott, J. Phys.: Condens. Matter \textbf{14}, 783 (2002).

\bibitem {r39} D.W. Brenner, Phys. Rev. B \textbf{42}, 9458 (1990).

\bibitem {SaitoBook} R. Saito, G. Dresselhaus, and M.S. Dresselhaus, \emph{Physical Properties of Carbon Nanotubes} (London: Imperial College Press, 1998).

\bibitem {Saito1998} R. Saito, T. Takeya, T. Kimura, G. Dresselhaus, and M.S. Dresselhaus, Phys. Rev. B \textbf{57}, 4145 (1998).

\bibitem {Wu2006} G. Wu and J. Dong, Phys. Rev. B \textbf{73}, 245414 (2006).

\bibitem {r40} A. Sears and R.C. Batra, Phys. Rev. B \textbf{69}, 235406 (2004).

\bibitem {r41} L. Yang, M. P. Anantram, J. Han, and J. P. Lu, Phys. Rev. B \textbf{60}, 13 874 (1999); L. Yang and J. Han, Phys. Rev. Lett. \textbf{85}, 154 (2000).

\bibitem {r42} B. I. Yakobson, C.J. Brabec, and J. Bernholc, Phys. Rev. Lett. \textbf{76}, 2511 (1996).

\bibitem {r43} M.B. Nardelli, B.I. Yakobson, and J. Bernholc, Phys. Rev. Lett. \textbf{81}, 4656 (1998).

\bibitem {WeakLink} K. R. Patton and M. R. Geller, Phys. Rev. B \textbf{64}, 155320 (2001); B. Hu, D.He, L. Yang and Y. Zhang, Phys. Rev. E \textbf{74}, R060101 (2006).

\bibitem {r44} B. Hu, L. Yang, and Y. Zhang, Phys. Rev. Lett. \textbf{97}, 124302 (2006).

\bibitem {r45} B.W. Smith, M. Monthioux, and D.E. Luzzi, Nature (London) \textbf{396}, 323 (1998).

\bibitem {r46} S. Okada, S.Saito, and A. Oshiyama, Phys. Rev. Lett. \textbf{86}, 3835 (2001).

\bibitem {r47} E.G. Noya, D. Srivastava, L.A. Chernozatonskii, and M. Menon, Phys. Rev. B \textbf{70}, 115416 (2004).

\end{thebibliography}
\end{document}